\documentclass{article}

\usepackage{PRIMEarxiv}

\usepackage[utf8]{inputenc} % allow utf-8 input
\usepackage[T1]{fontenc}    % use 8-bit T1 fonts
\usepackage{hyperref}       % hyperlinks
\usepackage{url}            % simple URL typesetting
\usepackage{booktabs}       % professional-quality tables
\usepackage{amsfonts}       % blackboard math symbols
\usepackage{nicefrac}       % compact symbols for 1/2, etc.
\usepackage{microtype}      % microtypography
\usepackage{lipsum}
\usepackage{fancyhdr}       % header
\usepackage{graphicx}       % graphics
\usepackage{multirow}
\usepackage{array}
\usepackage{amsmath}
\graphicspath{{media/}}     % organize your images and other figures under media/ folder

\title{FairTalk: Facilitating Balanced Participation in Video Conferencing by Implicit Visualization of Predicted Turn-Grabbing Intention
}

\author{
  Ryo Iijima \\
  OMRON SINICX Corporation\\
  \texttt{ryo.iijima@sinicx.com} \\
  \AND
  Shigeo Yoshida \\
  OMRON SINICX Corporation \\
  \texttt{shigeo.yoshida@sinicx.com} \\
  \And
  Atsushi Hashimoto \\
  OMRON SINICX Corporation \\
  \texttt{atsushi.hashimoto@sinicx.com} \\
  \And
  Jiaxin Ma \\
  OMRON SINICX Corporation \\
  \texttt{jiaxin.ma@sinicx.com} \\
}

\def\eqref#1{equation~\ref{#1}}
\def\1{\bm{1}}

\DeclareMathAlphabet{\mathsfit}{\encodingdefault}{\sfdefault}{m}{sl}
\SetMathAlphabet{\mathsfit}{bold}{\encodingdefault}{\sfdefault}{bx}{n}

\def\gX{{\mathcal{X}}}

\newcommand{\R}{\mathbb{R}}
\newcommand{\N}{\mathbb{N}}

\begin{document}
\maketitle

\begin{abstract}
Creating fair opportunities for all participants to contribute is a notable challenge in video conferencing.
This paper introduces \textit{FairTalk}, a system that facilitates the subconscious redistribution of speaking opportunities.
FairTalk predicts participants' turn-grabbing intentions using a machine learning model trained on web-collected videoconference data with positive-unlabeled learning, where turn-taking detection provides automatic positive labels. To subtly balance speaking turns, the system visualizes predicted intentions by mimicking natural human behaviors associated with the desire to speak.
A user study suggests that FairTalk may help improve speaking balance, though subjective feedback indicates no significant perceived impact. 
We also discuss design implications derived from participant interviews.
\end{abstract}

\begin{figure}[h]
  \includegraphics[width=\textwidth]{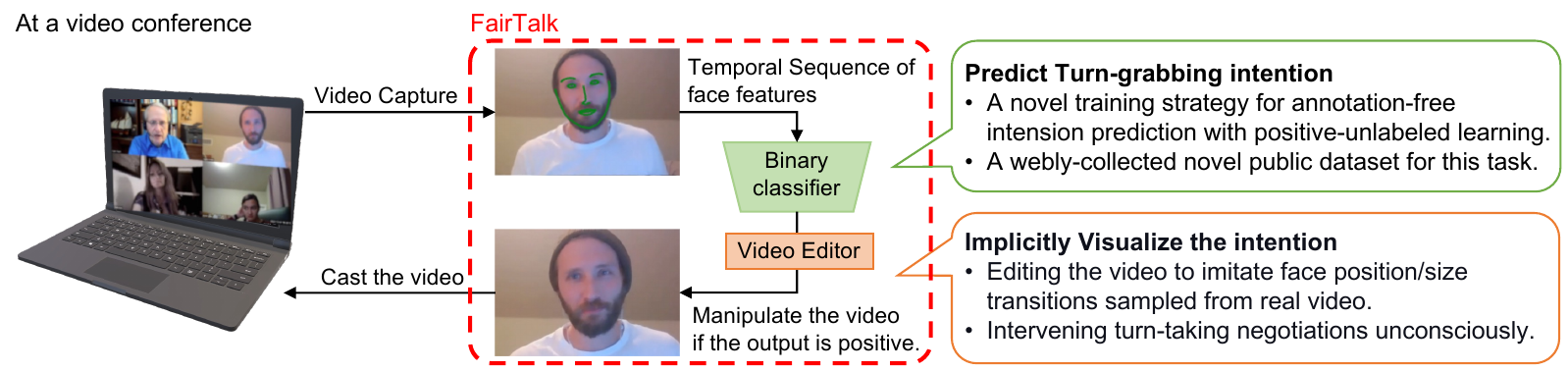}
  \caption{We designed a system, ~\textit{FairTalk}, that automatically facilitates video conferencing to promote a fair talking opportunity distribution.
  It predicts listeners' intention of turn-grabbing (willingness to speak) by a machine learning model trained on a novel unsupervised framework based on positive-unlabeled (PU) learning.
  %; positive samples are identified automatically via speaker change detection, and others are treated as unlabeled (i.e., negative or potentially positive).
  An implicit visualization with pseudo-action helps participants unconsciously balance the talking opportunity.
  (Photos from \url{https://vimeo.com/654390624})
  }
  \label{fig:teaser}
\end{figure}

\section{Introduction}

Video conferencing has become a cornerstone of modern communication, transcending physical boundaries to enable collaboration across the globe. However, fostering balanced participation remains a persistent challenge, as assertive participants often dominate discussions while others struggle to contribute, hindering the equitable exchange of ideas and limiting meeting effectiveness.

Previous works have primarily tackled participation imbalances using explicit visualizations of engagement metrics or speaking turn statistics, often providing feedback during or after meetings~\cite{TableTalk_Ogawa12, li2022conversation, samrose2018CoCo}. While these approaches encourage conscious behavioral adjustments, they impose cognitive loads that disrupt the natural flow of conversation. Inspired by the mindless computing paradigm, which subtly influences behavior without requiring user awareness~\cite{MindlessComputing_Adams15}, we explore a more seamless and intuitive alternative: implicit facilitation.

In this study, we introduce \emph{FairTalk}, a novel system that promotes balanced participation in video conferencing. FairTalk employs a machine learning framework for predicting turn-grabbing intentions—participants' readiness to speak—without manual annotations by leveraging positive-unlabeled (PU) learning~\cite{LearningTo_Li03,uPU,nnPU}, a technique that trains models using clear positive examples and unlabeled data. The system subtly visualizes predictions using a \textit{pseudo-leaning-forward effect}, mimicking natural preparatory actions~\cite{Reducing_Tamaki11} to foster balanced interactions unconsciously, maintaining conversational flow~\cite{nakazato2014smart,suzuki2017faceshare}.

Our contributions are fourfold: 
\begin{itemize} 
\item We propose an unsupervised framework for turn-grabbing intention prediction using PU learning and turn-taking detection. 
\item We curate a public dataset\footnote{The dataset and code is available at \url{https://github.com/omron-sinicx/FairTalk} for reproducible research, including 135 hours of training data and fully labeled test and validation sets.}
\item We design an implicit interaction mechanism that subtly influences behavior, avoiding cognitive overload. 
\item We explore the system’s potential to promote balanced speaking opportunities through user studies, offering insights for future design. 
\end{itemize}

By integrating mindless computing principles into virtual communication, FairTalk provides a scalable, user-friendly solution to enhance equity in participation.

\section{Related Work}

\subsection{Machine Learning for Turn-Taking}
Turn-taking is a core component of human communication, where the right to speak shifts between participants~\cite{SomeFunction_Kendon67, ASimplest_Sacks78}. Early works investigated its structure and rules~\cite{TurnTaking_Levinson16}. Machine learning approaches have modeled turn-taking behaviors by leveraging verbal cues like acoustic and linguistic data~\cite{IsTheSpeaker_Ferrer02, NeuralDialogue_Masumura18, Exploring_Skantze15, MultiMediate_Muller21, TA-CNN_Ma22}. However, these methods primarily target speaker actions, limiting their effectiveness for listener-driven turn-grabbing predictions.

Non-verbal features, such as gaze~\cite{PredictionOf_Kawahara12, GazeAnd_Jokinen13}, head movements~\cite{PredictingNext_Ishii15}, and respiration~\cite{UsingRespiration_Ishii16}, offer more relevance for listener actions. While some approaches integrate verbal and non-verbal signals~\cite{MultimodalFloor_Chen09, TrimodalPrediction_Ishii22}, they often rely on supervised learning with annotated datasets, posing scalability challenges. Unlike these methods, we propose an unsupervised PU learning framework~\cite{LearningTo_Li03, nnPU} to predict turn-grabbing intentions, where turn-taking detection provides automatically labeled positive samples, reducing reliance on manual labeling.

\subsection{Facilitation Systems for Balanced Participation}
Several facilitation systems have visualized engagement metrics or interaction summaries to encourage balanced participation~\cite{ConversationClock_Bergstrom07, TableTalk_Ogawa12}. These mindful computing approaches rely on explicit feedback to promote self-regulation among participants~\cite{AnInteractive_Bachour10, Influencing_DiMicco04}. In contrast, FairTalk draws on mindless computing principles~\cite{MindlessComputing_Adams15}, implicitly influencing behavior without conscious user effort.

Video conferencing presents unique challenges in turn negotiation, as speaker interruptions occur less frequently than in face-to-face conversations~\cite{RemoteConversations_Sellen95}, particularly with larger participant groups~\cite{VideoConferencing_Gowan94}. FairTalk addresses this by subtly guiding attention toward listeners likely to speak, promoting more balanced exchanges without explicit intervention.

\subsection{Interaction-Supporting Technologies for Video Conferencing}
Visualization tools linking gaze and turn-taking have enhanced remote communication~\cite{EyeView_Jenkin05, GazeChat_He21}. Systems such as MirrorSpace~\cite{MirrorSpace_Roussel04} and FluidMeet~\cite{FluidMeet_Hu22} used virtual proximity to support interaction, while OpenMic~\cite{OpenMic_Hu23} introduced spatial organization for managing turns. However, these systems typically require mindful user participation.

Explicit visualization methods for smoother turn-taking, such as motion detection~\cite{Reducing_Tamaki11}, prompt users to perform predefined gestures, creating cognitive demands. FairTalk extends this concept by adopting implicit visualization with pseudo-actions~\cite{nakazato2014smart, suzuki2017faceshare}, promoting balanced participation by mimicking natural preparatory behaviors without explicit user effort.

\renewcommand{\arraystretch}{1.2}
\section{Design and Implementation}

The FairTalk system is designed to facilitate balanced participation in video conferencing by predicting participants’ turn-grabbing intentions and visualizing these predictions implicitly. This section outlines the technical foundations, data processing pipeline, and system architecture.

\subsection{Positive-Unlabeled Learning}

We designed a classifier to determine whether a participant demonstrates a turn-grabbing intention (positive) or not (negative) based on observable behavior. While the traditional method involves manually labeling each instance and applying positive-negative (PN) learning, this approach is inefficient and unsuitable for large-scale datasets due to its dependence on human annotation.

To address this limitation, we adopt positive-unlabeled (PU) learning~\cite{BuildingText_Liu03,uPU, nnPU}, a technique that uses a small set of labeled positive samples while treating the rest of the data as unlabeled, potentially containing both positive and negative instances.

In video conferencing, turn-grabbing intentions are latent and must be inferred from participant behavior. In our study, turn-taking events serve as observable signals for identifying positive turn-grabbing intentions. Specifically, when a participant becomes a new speaker, it implies they are likely to have a turn-grabbing intention immediately before the previous speaker’s utterance ends. This allows us to automatically extract positive samples by selecting short video clips preceding turn-taking events. All other data, which may contain both positive and negative instances, are treated as unlabeled. The detailed problem formulation of applying PU learning is given in \autoref{appx:formulation}.

\subsection{Data Processing and Model Training}
\begin{figure}[ht]
  \centering
  \includegraphics[width=\linewidth]{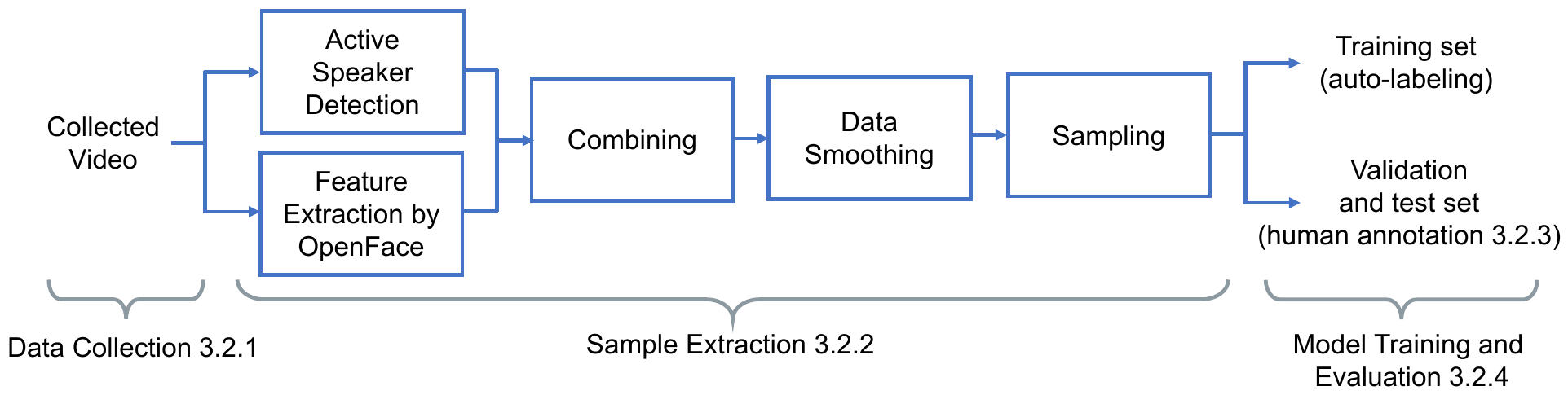}
  \caption{The flowchart of data processing.}
  % \Description{}
  \label{fig:data_process}
\end{figure}

Our data processing pipeline (\autoref{fig:data_process}) spans from collecting raw video data to generating positive and unlabeled samples for PU learning, followed by dataset splits for training, validation, and testing. Positive and negative labels were manually added to the validation and test sets for evaluation.

All raw data were sourced from publicly available platforms, ensuring accessibility for replication. We also commit to sharing our processed dataset to support future research.

\subsubsection{Data Collection}
We gathered 121 video recordings from YouTube and Vimeo, focusing on conversations involving 3 to 9 participants, predominantly in gallery view, and primarily in English. This dataset includes 112 recordings (135.8 hours) for training, 3 recordings (2.8 hours) for validation, and 6 recordings (4.2 hours) for testing.

\subsubsection{Sample Extraction}
To detect turn-taking events, we applied Active Speaker Detection (ASD)~\cite{tao2021someone} alongside OpenFace~\cite{openface_2018} for facial feature extraction. A smoothing operation was applied to mitigate ASD glitches. A four-second interval, starting up to 10 seconds before speech onset, was randomly sampled as input data. Refer to the \autoref{appx:preprocessing} for a detailed description.

\subsubsection{Human Annotation} \label{sssec:human}
While PU learning does not require negative samples for training, model evaluation necessitates fully labeled data. We manually labeled 210 validation and 420 test samples (70 samples each from 9 videos). Annotators classified four-second clips cropped to facial regions into one of the following four categories:
\begin{enumerate}
\item Intends to remain silent (negative).
\item Willing to speak if prompted (possibly positive).
\item Intends to speak spontaneously (positive).
\item Already speaking (outlier).
\end{enumerate}
The outlier category addressed ASD inaccuracies and was excluded from the evaluation. 
Annotation results (\autoref{tb:annotation}) reveal a class imbalance favoring negative samples and highlight greater disagreement among annotators in the test set compared to the validation set, suggesting higher difficulty in the former.

\begin{table*}[h]
\caption{Annotation results where values represent sample counts. The Majority Voting (MV) results among the three annotators were used as final labels.} \label{tb:annotation}
% \begin{tabular}{lC{1.2cm}C{1.2cm}C{1.2cm}C{1.2cm}C{1.2cm}C{1.2cm}C{1.2cm}C{1.2cm}} \toprule
\begin{tabular}{l>{\centering\arraybackslash}m{1.2cm}>{\centering\arraybackslash}m{1.2cm}>{\centering\arraybackslash}m{1.2cm}>{\centering\arraybackslash}m{1.2cm}>{\centering\arraybackslash}m{1.2cm}>{\centering\arraybackslash}m{1.2cm}>{\centering\arraybackslash}m{1.2cm}>{\centering\arraybackslash}m{1.2cm}} 

\multirow{2}{*}{} & \multicolumn{4}{c}{Validation Set} & \multicolumn{4}{c}{Test Set} \\ 
                    \cmidrule(lr){2-5} \cmidrule(lr){6-9}
                  & Annot. 1 & Annot. 2 & Annot. 3 & MV & Annot. 1 & Annot. 2 & Annot. 3 & MV \\  \midrule
Negative          & 100 & 124 & 121 & 123 & 331 & 189 & 215 & 275 \\
Poss. Positive & 71  & 46  & 63  & 40  & 39  & 90  & 156 & 59  \\
Positive          & 39  & 40  & 26  & 32  & 22  & 114 & 19  & 18  \\
Outlier           & 0   & 0   & 0   & 0   & 28  & 27  & 30  & 35  \\
Total             & 210 & 210 & 210 & 195 & 420 & 420 & 420 & 387 \\  \bottomrule 
\end{tabular}
\end{table*}

\subsubsection{Model Training and Evaluation}
We selected 19 features from OpenFace, focusing on 17 Facial Action Units and 2 Gaze directions, excluding other features to reduce overfitting observed in preliminary tests. Our lightweight network architecture consists of two 1D convolutional layers followed by an LSTM and a final fully connected layer for binary classification. Refer to \autoref{appx:hyperparameter} for detailed model architecture and hyperparameter tuning.

Using the model mentioned above, we achieved a Matthews Correlation Coefficient (MCC) of 0.175, an Area Under the Curve (AUC) of 0.602, an F-score of 0.392, an accuracy of 0.648, a precision of 0.315, and a recall of 0.519 on the test set. The moderate values of MCC and AUC highlight the inherent difficulty of recognizing turn-grabbing intentions, a task that involves predicting subtle, context-dependent user behaviors. This complexity leads to inevitable false positives and false negatives in practical use. To minimize their impact on user experience, we adopted the implicit visualization design described in the following section, which mitigates the influence of prediction inaccuracies.

\subsection{Implicit Feedback Design}

\subsubsection{Design Considerations}
Our design for subtle, mindless facilitation of turn-grabbing intentions during video conferencing follows two key principles derived from prior research~\cite{MindlessComputing_Adams15}:
\begin{enumerate}
\item Leverage unconscious behaviors: People naturally perform certain behaviors, such as leaning forward, to signal turn-grabbing intentions without conscious effort. Mimicking these behaviors avoids relying on artificial visual symbols, enhancing intuitive communication.
\item Use standard devices: To ensure accessibility and ease of use, the system should function with commonly available videoconferencing hardware.
\end{enumerate}
Based on these criteria, we modeled the ``leaning forward'' behavior, a preparatory action signaling an intent to speak~\cite{Reducing_Tamaki11}. While nodding or other gestures could be viable alternatives, real-time replication of such actions posed technical challenges. Therefore, we focused exclusively on a visual effect we call pseudo leaning-forward.

\subsubsection{Pseudo Leaning-Forward Effect}
Our goal was to replicate the leaning-forward effect as naturally as possible. Using data from our video recordings, we analyzed instances where participants leaned forward before speaking. OpenFace tracked changes in facial size and movement (both vertical and horizontal) over time. The system applies these recorded magnifications and shifts using frame-by-frame affine transformations, simulating a natural leaning-forward motion to convey turn-grabbing intention. The pseudo-leaning-forward effect subtly influences conversation flow without imposing cognitive load or making prediction errors overly conspicuous, thereby enhancing the usability and acceptance of the system despite the prediction challenges.

\subsection{System Pipeline}
The FairTalk system integrates real-time turn-grabbing prediction and implicit feedback delivery. It captures video from a webcam, processes the images to predict turn-grabbing intentions, and streams the augmented video through a virtual camera interface.

The process begins with OpenFace extracting facial features from the live video feed. These features, combined with historical data, form time-series input for the machine learning model, which outputs a prediction of turn-grabbing intention. If an intention is detected, the system applies the pseudo leaning-forward effect. The enhanced video stream is then transmitted using Open Broadcaster Software (OBS)\footnote{\url{https://obsproject.com}} for seamless integration into videoconferencing platforms.

\section{User study}

We conducted a controlled user study to evaluate the effectiveness of FairTalk, comparing implicit facilitation with baseline and explicit visualization conditions. The study tested two hypotheses:
\begin{itemize}
    \item H1: Visualizing predicted turn-grabbing intentions facilitates balanced speaking opportunities.
    \item H2: Implicit visualization is preferred over explicit methods for turn-grabbing intention presentation.
\end{itemize}

\subsection{Participants and Experimental Protocols}
The study involved 47 participants (25 male, 22 female, mean age = 24.6), grouped into twelve sessions of four participants each (one group had three participants due to an absence). Participants were assigned to one of three conditions:
(1) \textbf{Baseline} video conferencing without intervention,
(2) \textbf{Explicit} visualization using a yellow frame around detected speakers (same as ~\cite{Reducing_Tamaki11}), and
(3) \textbf{Implicit} visualization via the pseudo-leaning-forward effect.

The experiment was conducted in a quiet room using laptops, webcams, and wireless headsets with active noise cancellation. Participants sat individually with visual separation and green screen virtual backgrounds to minimize distractions and bias. They engaged in a video conferencing task comprising a practice session and a main session, with casual discussion topics requiring no specialized knowledge. Balanced group assignments were predetermined.

An initial instruction introduced the experiment and the application of one of three settings: baseline, explicit, or implicit. Participants viewed examples of turn-grabbing visual effects and were informed that following the effects was optional. The 10-min practice session used the topic ``\textit{Cats or dogs: Which one would you prefer at home?}'' and had no system intervention. After a 5-min break, the 15-min main discussion addressed one of two counter-balanced topics: ``\textit{Premade tour package or personalized travel plan?}'' or ``\textit{Paper books vs. e-books?}'' Participants then completed a questionnaire, provided demographic data, and joined a brief semi-structured interview. The total experiment duration was approximately 60 min.

\subsection{Results}

\subsubsection{Objective measurements}
We assessed balance in speaking opportunities using two metrics: (1) \textbf{Ratio of Speaking Turns (RSTurn)}, representing each participant's turns divided by the total, and (2) \textbf{Ratio of Speaking Time (RSTime)}, measuring individual speaking duration as a proportion of the total discussion time. The ideal value for both is 1/$N$, where $N$ is the number of participants. Short backchannels within one second were excluded from turn counts~\cite{GazeAnd_Jokinen13}.

\begin{figure*}[ht]
  \centering
  \includegraphics[width=\linewidth]{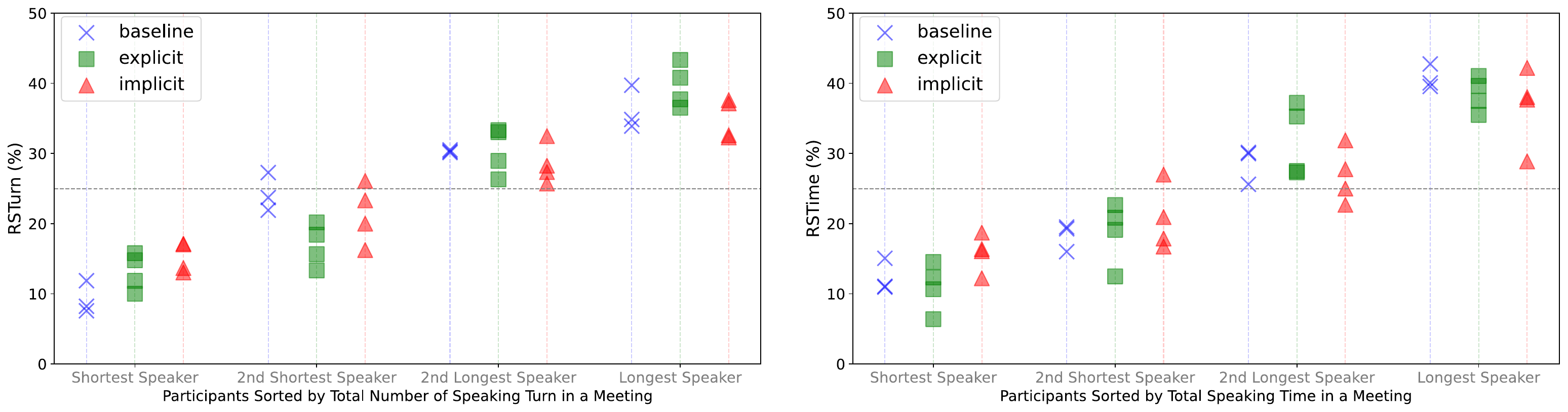}
  \caption{The distribution of speaking turns (left) and speaking time (right). Participants are ordered within each group from the shortest to the longest speaker for each metric.}
  % \Description{A scatter plot of speaking turns and the number of speaking times.}
  \label{fig:RSTurn_RSTime}
\end{figure*}

Comparison of RSTurn and RSTime across conditions is shown in \autoref{fig:RSTurn_RSTime}. Participants were ranked by speaking turns or speaking time within each group and labeled as the shortest, second shortest, second longest, or longest speaker. In the implicit condition, the lowest speaking turn ratio was higher, and time distribution was more balanced than in the baseline. Absolute differences from ideal ratios (25\%) also indicated smaller imbalances in the implicit setting. However, differences were marginal between explicit and implicit conditions.

\subsubsection{Subjective measurements}

Subjective measures used 7-point Likert scale questions adapted from prior studies on video conferencing and turn-taking~\cite{RemoteConversations_Sellen95}. Participants rated their experiences in categories such as ease of expressing themselves, conversation control, interruptions, pauses, and overall satisfaction.

\begin{figure}[ht]
  \centering
  \includegraphics[width=\linewidth]{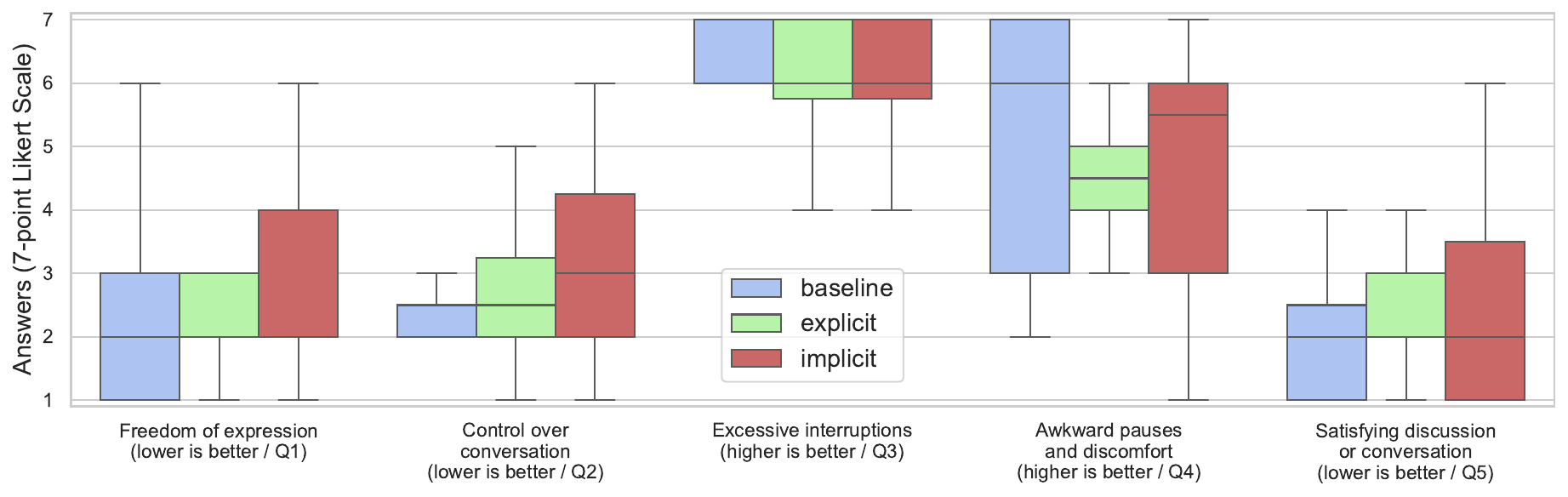}
  \caption{The distribution of responses for each Likert question item (1: strongly agree; 7: strongly disagree).}
  % \Description{A strip plot of the Likert question items.}
  \label{fig:likert}
\end{figure}

Subjective evaluations (\autoref{fig:likert}) revealed no significant differences between conditions across survey items (p-values ranged from 0.35 to 0.61). 

\subsubsection{Interview}
In the interview, no participants missed the effect in the explicit condition, but six participants missed it in the implicit condition. Regarding false system responses, six noticed false positives in the explicit condition and three in the implicit. For false negatives, one participant in each condition expressed disappointment when desired effects did not occur, while two in the implicit condition recognized the system's correct responses at appropriate moments, expressing satisfaction.

Several participants felt encouraged to contribute after noticing the effect, but none used it to control the conversation. One noted a lack of confidence in interrupting to redistribute the floor. Two participants appreciated the subtlety of the implicit condition, fearing more noticeable effects would distract or make speakers anxious.

The experiment inspired ideas for improvement, with one participant suggesting the system would be more beneficial in larger groups where not everyone speaks regularly, and another thinking it could help in intense debates by intervening when participants have strong opposing views.

\subsection{Discussion and Design Implications}

The results suggest that implicit visualization of speech intentions may influence conversation dynamics. 
In the implicit condition, the least talkative person tended to speak more, and the most talkative person spoke slightly less, indicating a possible effect where visualizing speech intentions encourages quieter participants to speak and prompts dominant individuals to yield the floor. 
In contrast, in the baseline condition, the least talkative person showed a tendency to speak less, and the explicit condition often led to more dominant participants due to overly prominent turn-grabbing cues. 
These findings imply that implicit visualizations might help regulate speaking durations by subtly influencing how time is allocated.

However, no significant differences were found in Likert scale responses across conditions, possibly because users adapted quickly to the systems, and changes in behavior were not consciously perceived. Some participants reported initially noticing the effects but gradually ignoring them, supporting this idea.

From a design perspective, it is crucial to treat the display of turn-grabbing intentions to the speaker and other participants as separate systems. Participants noted that seeing their own intentions triggered a desire to speak next, but it could also lead to undue pressure. In contrast, showing the effects to others might foster a more natural flow. Implicit visualization, being subtler, avoids the potential for increased distrust that might arise from inaccurate explicit cues.

Finally, the study highlights the importance of subtly influencing conversation flow. The implicit effect did not force turn-taking but gently encouraged shifts in speaking turns, aligning with the goal of promoting natural, face-to-face-like communication. This suggests that fostering awareness of turn-grabbing intentions in a subtle way is more effective than overt disruptions to the conversation.

\section{Limitation and Future Work}

This study highlights promising results for FairTalk, yet several limitations must be addressed in future research. One concern is the cultural and demographic constraints of the user study, which exclusively involved Japanese participants, while the dataset used for training was derived from Western video conferences. This cultural mismatch may have influenced both model performance and participant behavior, suggesting the need for cross-cultural validations and model tailoring for diverse demographics. Another challenge is the generalizability of the prediction model. Although PU learning reduced annotation costs, the model’s performance metrics (MCC = 0.175, AUC = 0.602) indicate room for improvement. Enhancements through domain generalization techniques~\cite{li2018domain,Li2018MLDG} or fine-tuning for specific user groups could improve accuracy.

Additionally, the study’s evaluation relied on objective metrics such as speaking turns and time, which may not fully capture cognitive load or subjective user experiences. Future studies might benefit from integrating diverse metrics like SPCCS~\cite{mccroskey1988self}, PRCA-24~\cite{mccroskey2015introduction}, and the WTC scale~\cite{mccroskey1992reliability}, to understand varied participant responses based on their attributes and abilities. Finally, the sample size and short interaction durations may limit the generalizability of the findings. Larger-scale, longer-term online experiments could provide more comprehensive insights and validate FairTalk’s applicability in real-world settings.

\section{Conclusion}

This paper introduced FairTalk, a system that facilitates balanced participation in video conferencing by predicting turn-grabbing intentions and implicitly visualizing them using a pseudo-leaning-forward effect. FairTalk leverages a scalable, annotation-efficient learning framework for intention prediction and an intuitive visualization design that fosters equitable communication without imposing cognitive burdens. A curated dataset is provided to support reproducibility in this domain. 
Insights from the user study suggest that the system may support improvements in speaking time and turn-taking balance. 
Specifically, it appeared to reduce the dominance of talkative participants and increase opportunities for quieter ones, aligning with the goal of promoting more inclusive conversations. 
The results indicate FairTalk’s potential to contribute participation equity in collaborative and competitive communication scenarios.

\section*{Acknowledgments}
We thank Maya Torii for helping us with the user study. We appreciate all the people who joined the pilot study and provided us insightful feedbacks.

%Bibliography
\bibliographystyle{unsrt}  
\bibliography{references}

\appendix
\section{Problem Formulation} \label{appx:formulation}

We implement PU learning with the following training sample configuration:
\begin{itemize}
    \item Positive: Data just before the acquisition of turns. They cover a part of the positive examples.
    \item Unlabeled: Data other than positive ones. They include positive examples that listeners did not put their intention into action and all negative examples.
\end{itemize}
Let $X \in \R^d$ ($d \in \N$) be an input and $Y \in \{-1, +1\}$ be an output of random variables, assuming an unknown distribution $P(X, Y)$.
When constructing the dataset with the problem setting mentioned above, positive (P) data of size $n_p$ would be sampled from $P(x | Y=+1)$ and unlabeled (U) data of size $n_u$ would be sampled from $P(x)$.
That is, $\gX_p = \{x^p_i\}^{n_p}_{i=1} \sim P(x|Y=+1)$ and $\gX_u = \{x^u_i\}^{n_u}_{i=1} \sim P(x)$.
From these two datasets, $\gX_p$ and $\gX_u$, the goal of PU learning is to train a classifier $f: \R^d \rightarrow [0, 1]$ that correctly classifies $x$ (with a thresholding the output).

In PN learning, we can access $\gX_n = \{x^p_i\}^{n_n}_{i=1} \sim P(x|Y=-1)$, where $n_n$ is the size of negative data.
The learning process minimizes the risk of prediction results being false negative or false positive.
This is implemented by estimating the expected loss $R_{pn}(f)$ from the empirical loss $\hat{R}_{pn}(f)$, with the goal of finding an $f$ that minimizes $R_{pn}(f)$.
The $\hat{R}_{pn}(f)$ can be computed as follows:
$$
\hat{R}_{pn} = \pi \hat{R}_p(f) + (1-\pi)\hat{R}_n(f)
$$
where
$$
\hat{R}_p(f) = \frac{1}{n_p}\sum_{x_i \in \gX_p} \ell(f(x_i), +1),\ \
\hat{R}_n(f) = \frac{1}{n_n}\sum_{x_i \in \gX_n} \ell(f(x_i), +1),\ \ 
\pi = P(Y=+1)
$$
and $\ell(t, y)$ represents the loss function where $t$ is the predicted value and $y$ is the ground truth.

On the other hand, in PU learning, we cannot access to $\gX_n$ at training.
To approximate PN learning, $\hat{R}_n(f)$ is replaced by another function that can be computed from $\gX_p$ and $\gX_u$.
While various methods have been proposed to do this ~\cite{LearningTo_Li03,BuildingText_Liu03,uPU,nnPU}, our study is the first to use PU learning to estimate intentions from events in human dialogue.
Therefore, we avoided heuristic methods that might only work in certain applications and adopted the Non-Negative Risk Estimator (referred to as nnPU) ~\cite{nnPU}. It is a state-of-the-art method with a theoretical guarantee, and we can expect that it works well regardless of dataset biases.

The nnPU method ~\cite{nnPU} aims to minimize the empirical loss $\hat{R}_{\text{nnpu}}(g)$ described by the following:
$$
\hat{R}_{\text{nnpu}}(g) = \pi \hat{R}_p^{+} (g) + \max\Big\{0, \hat{R}_u^- (g) - \pi \hat{R}_p^- (g)\Big\}
$$
where
\begin{equation*}
\begin{split}
\hat{R}_p^+ (g) &= \frac{1}{n_p}\sum_{x_i \in \gX_p} \ell(g(x_i), +1),\\
\hat{R}_p^- (g) &= \frac{1}{n_p}\sum_{x_i \in \gX_p} \ell(g(x_i), -1),\\
\hat{R}_u^- (g) &= \frac{1}{n_u}\sum_{x_i \in \gX_p} \ell(g(x_i), -1)
\end{split}
\end{equation*}

\section{Data Preprocessing} \label{appx:preprocessing}
We conducted the following preprocessing steps on the downloaded videos:
\begin{itemize}
    \item We used two methods, OpenFace~\cite{openface_2018} for feature extraction and TalkNet~\cite{tao2021someone}%\footnote{\url{https://github.com/TaoRuijie/TalkNet-ASD}} 
    for Active Speaker Detection (ASD). These methods identify human faces in the videos, providing face IDs and temporal information, including bounding boxes, facial feature values, and ASD scores indicating the likelihood of being a speaker.
    \item To synchronize data, we matched bounding boxes from both OpenFace and ASD results for each timestamp. When the bounding boxes intersected, we combined OpenFace features with speaker information from ASD. In cases where OpenFace features were missing (typically due to face detection failures), we performed interpolation to fill the gaps.
    \item To improve data quality, we applied a data smoothing operation to the ASD scores for each face ID. This operation helps reduce the impact of glitches. Specifically, if there was an interval marked as ``speaking'' by ASD that lasted less than 1 second, we reclassified it as ``non-speaking'', and vice versa.
\end{itemize}
After completing the above preprocessing steps, we move on to the data sampling process.

\begin{figure}[ht]
  \centering
  \includegraphics[width=\linewidth]{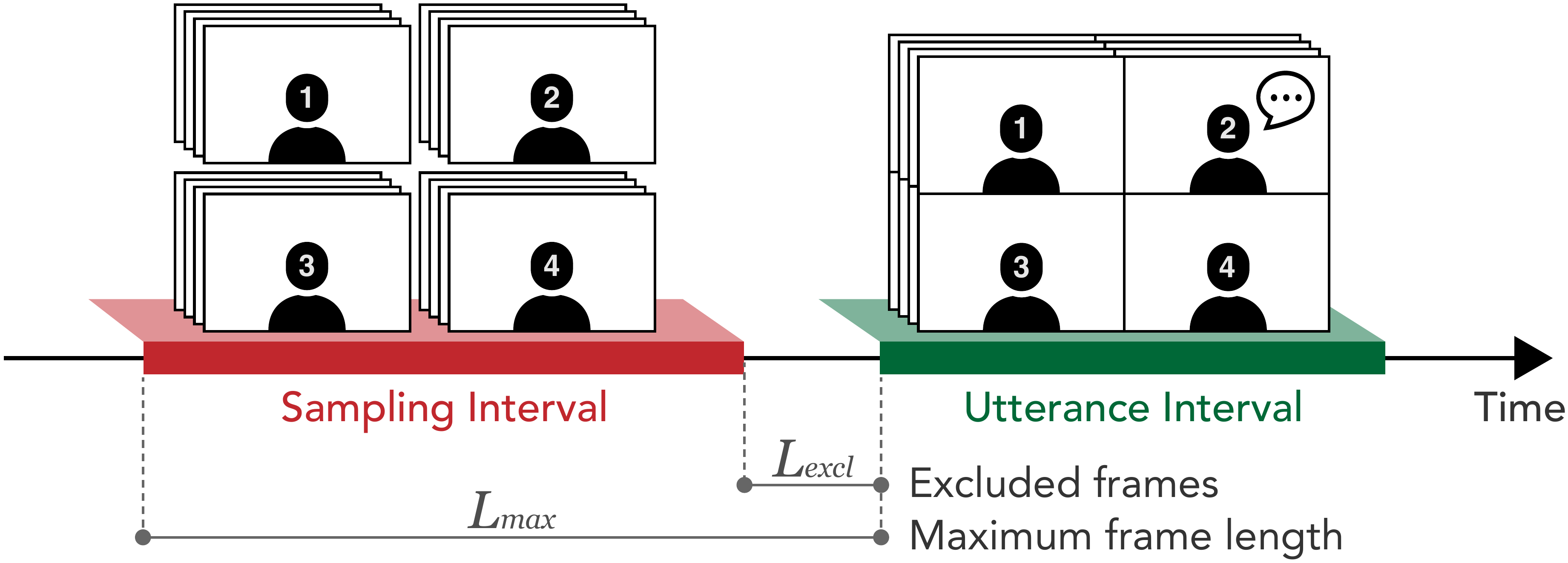}
  \caption{An example of face feature sequence sampling.}
  \label{fig:sample_extraction}
\end{figure}

Our sampling procedure is illustrated in ~\autoref{fig:sample_extraction}.
A random four-second interval was sampled from the time period starting from $L_{max}$ prior to the onset of speech up to the point $L_{excl}$ (representing frames of potential mouth-opening) before the same moment.
The value for $L_{max}$ was set to 10 seconds.

\section{Model Architecture and Hyperparameter Tuning} \label{appx:hyperparameter}

\begin{figure}[ht]
  \centering
  \includegraphics[width=0.8\linewidth]{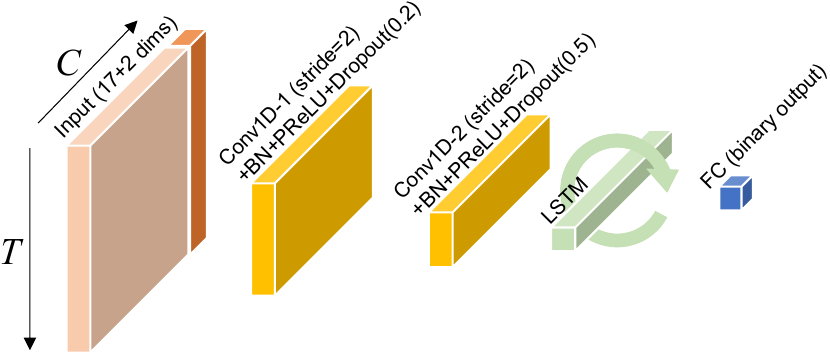}
  \caption{Network architecture for turn-grabbing intention prediction. $T$ and $C$ are the length of time series data and the feature dimensionality, respectively. Conv1D is applied in the temporal direction.}
  \label{fig:network_architecture}
  % \Description{A 19-dim input is fed to two 1D convolutional layers, then LSTM and fully-connected layers. The output is a binary classification result.}
\end{figure}

Due to the possible overfitting, we designed the network architecture as simply as possible, as shown in \autoref{fig:network_architecture}. Its first two layers are 1-dimensional convolutional layers (Conv1Ds), known as a classical but promised parameter-efficient structure for time series data processing.
Long Short-Term Memory (LSTM) follows the Conv1Ds to grab the unknown temporal event correlations in the time series.
The output of LSTM is finalized to a binary logit with a single fully connected layer (FC). 
We tuned hyperparameters with the help of the human-annotated validation set.

For hyperparameter tuning, we employ Optuna ~\cite{optuna_2019}, an automated hyperparameter optimization framework, to conduct a grid search. Our optimization target is the Matthews Correlation Coefficient (MCC) on the validation set. As mentioned in \autoref{sssec:human}, since the validation set contains three classes while the training set consists of binary classes, through performance comparison, we categorized the ``Possibly Positive'' class as ``Positive'' during validation.

For each iteration, we run 50 epochs. After the first five steps, we implement median pruning, which terminates trials that yield worse results than the median of previous trials. 

The ranges of hyperparameters explored include:
\begin{itemize}
    \item Conv1D 1st-layer dimension: $\{\underline{8}, 16, 32, 64\}$
    \item Conv1D 2nd-layer dimension: $\{8, 16, 32, 64, \underline{128}\}$
    \item LSTM number of layers: $\{1, \underline{2}, 3, 4, 5, 6\}$
    \item LSTM layers dimension: $\{\underline{16}, 32, 64, 128\}$
    \item Learning rate: $\{\underline{1\mathrm{e}{-2}}, 1\mathrm{e}{-3}, 1\mathrm{e}{-4}, 1\mathrm{e}{-5}\}$,
\end{itemize}
where underline numbers correspond to the optimal hyperparameter combination found by Optuna.

\end{document}